\documentclass[%
 reprint,
 amsmath,amssymb,
 aps,
prb,
]{revtex4-2}

\usepackage[utf8]{inputenc}
\usepackage{graphicx}
\usepackage{overpic}
 \usepackage{rotating}
 \usepackage{mwe, tikz}
 \usepackage{amsmath,amssymb}
\usepackage{pgfplots}
\usetikzlibrary{calc,arrows,decorations.markings,shapes.geometric} 
\usepackage{amssymb}
\usepackage[finalnew]{trackchanges}
\usepackage{lipsum}
\usepackage{romannum}
\soulregister\ref7
\soulregister\cite7
\soulregister\citeA7
\soulregister\romannum7

\usepackage{graphicx}
\usepackage{amsmath}
\usepackage{mathtools} 
\usepackage{color}
\usepackage{overpic}
\usepackage[export]{adjustbox}
\usepackage{mwe, tikz}
\usepackage{rotating}
\usepackage{latexsym}
\usepackage{csquotes}
\usepackage{tikz} 
\usetikzlibrary{calc,arrows,decorations.markings} 
\usepackage{pict2e}
\usepackage{pgfplots}
\usepackage{float}
\usepackage{hyperref}
\usepackage{xr}
\usepackage{rotating}
\usepackage{amsmath,amssymb}
\usepackage{fp}
\usetikzlibrary{calc} 
\usepackage{booktabs, tabularx}
\usepackage{romannum}

\newcommand{\legSqTxt}[3]{
		\begin{tikzpicture}
			     \coordinate (center1) at (#1,#2); 
                \coordinate (b) at ($ (center1) - (.075,0) $);
                \coordinate (c) at ($ (b) - (.1,0) $); 
                \coordinate (d) at ($ (center1) + (.075,0) $);
                \coordinate (e) at ($ (d) + (.1,0) $); 
				 \draw[black,fill=#3] ($(b)-(0.0,0.075)$) rectangle ($ (d) + (0.0,0.075) $); 
		\end{tikzpicture}		
}

\newcommand{\circTxtFill}[3]{
		\begin{tikzpicture}
                \coordinate (center1) at (#1,#2); 
				 \draw[black,fill=#3] (center1) circle (2pt);
		\end{tikzpicture}		
}

\newcommand{\legDiamondTxt}[3]{
		\begin{tikzpicture}
			     \coordinate (center1) at (#1,#2); 
                \coordinate (b) at ($ (center1) + (0.075,0.0) $); 
                \coordinate (c) at ($ (center1) + (0,0.075) $); 
                \coordinate (d) at ($ (center1) + (-0.075,0) $);
                \coordinate (e) at ($ (center1) + (0.0,-0.075) $);
				 \draw[#3] (b) -- (c) -- (d) -- (e) -- (b);
                \coordinate (c) at ($ (b) + (.1,0) $); 
                \coordinate (e) at ($ (d) - (.1,0) $); 
		\end{tikzpicture}		
}
                
\newcommand{\legTriangleTxt}[3]{
		\begin{tikzpicture}
			     \coordinate (center1) at (#1,#2); 
                \coordinate (b) at ($ (center1) - 0.17*(0.5,0.289) $); 
                \coordinate (c) at ($(center1) + 0.17*(0.5,-0.289) $); 
                \coordinate (d) at ($ (center1) + 0.17*(0,0.366) $);
				\draw[black,fill=#3, line width=0.1mm] 
    (b) -- (c) -- (d) -- cycle;                 
    			\coordinate (b) at ($ (center1) - (.075,0) $);
                \coordinate (c) at ($ (b) - (.1,0) $); 
                \coordinate (d) at ($ (center1) + (.075,0) $);
                \coordinate (e) at ($ (d) + (.1,0) $); 
		\end{tikzpicture}		
}		
		
\definecolor{light-gray}{gray}{0.85}

\newcommand{\leftTriang}[3]{
		\begin{tikzpicture}
			     \coordinate (center1) at (#1,#2); 
                \coordinate (b) at ($ (center1) + 0.17*(0.289,0.5) $); 
                \coordinate (c) at ($ (center1) + 0.17*(0.289,-0.5) $); 
                \coordinate (d) at ($ (center1) + 0.17*(-0.366,0.0) $);
				 \draw[#3] (b) -- (c) -- (d) -- (b);
                \coordinate (b) at ($ (center1) - (.075,0) $);
                \coordinate (c) at ($ (b) - (.1,0) $); 
                \coordinate (d) at ($ (center1) + (.075,0) $);
                \coordinate (e) at ($ (d) + (.1,0) $); 
		\end{tikzpicture}		
}

\newcommand{\rightTriang}[3]{
		\begin{tikzpicture}
			    \coordinate (center1) at (#1,#2); 
                \coordinate (b) at ($ (center1) + 0.17*(-0.289,0.5) $); 
                \coordinate (c) at ($ (center1) + 0.17*(-0.289,-0.5) $); 
                \coordinate (d) at ($ (center1) + 0.17*(0.366,0.0) $);
				 \draw[black,fill=#3] (b) -- (c) -- (d) -- (b);
                \coordinate (b) at ($ (center1) - (.075,0) $);
                \coordinate (c) at ($ (b) - (.1,0) $); 
                \coordinate (d) at ($ (center1) + (.075,0) $);
                \coordinate (e) at ($ (d) + (.1,0) $); 
		\end{tikzpicture}		
 }
\newcommand{\drawSlope}[6]{ 
				\coordinate (center1) at (#1,#2); 
				 \FPeval{\nx}{cos(#4*pi/180)}%
				 \FPeval{\ny}{sin(#4*pi/180)}%
				 \FPeval{\absNx}{abs(\nx)}%
				 \FPeval{\absNy}{abs(\ny)}%
				\coordinate (b) at ($ (center1) + #3*(-\ny,+\nx) $);
				\coordinate (c) at ($ (center1) - #3*(-\ny,+\nx) $); 
				 \FPeval{\xx}{(-\nx*\ny)}%
				\ifdim\xx pt < 0pt 
				\coordinate (d) at ($ (c) -2*#3*\ny*(1,0) $);
				\node at ($(d) +(0,#3*\nx)-0.2*(\nx/\absNx,0)$) {{\color{#5}$#6$}}; 
				\node at ($(d)+(#3*\ny,0)-0.2*(0,\ny/\absNy)$) {{\color{#5}$\scriptstyle 1$}}; 
				\else
				\coordinate (d) at ($ (c) +2*#3*\nx*(0,1) $);
				\node at ($(d)-#3*\nx*(0,1)-0.2*\nx/\absNx*(1,0)$) {{\color{#5}$#6$}}; 
				\node at ($(d)-#3*\ny*(1,0)-0.2*\ny/\absNy*(0,1)$) {{\color{#5}$\scriptstyle 1$}}; 
				\fi
				\draw[#5,line width=0.1mm] (d) -- (b); 
				\draw[#5,line width=0.1mm] (d) -- (c); 
				\draw[#5,line width=0.1mm] (b) -- (c); 
} 

\newcommand{\scaleBarr}[8]{
			     \coordinate (center1) at (#1,#2); 
                \coordinate (b) at ($ (center1) - (#3,#4/2) $); 
                \coordinate (d) at ($ (center1) + (#3,#4/2) $);
				 \draw[black,fill] ($(b)$) rectangle ($(d)$); 

				\node at ($(b)-(0,#4)$){{\color{black} $#5$}};
				\node at ($(b)+(2*#3,-#4)$) {{\color{black} $ #6$}};
				\node at ($(b)+(2*#3,+2.6*#4)$) {{\color{black}$\text{#8}$}}; 
			     \coordinate (center2) at ($ (center1) + 2*(#3,0.0)$);  
                \coordinate (b) at ($ (center2) - (#3,#4/2) $);
                \coordinate (d) at ($ (center2) + (#3,#4/2) $);
				 \draw[black] ($(b)$) rectangle ($(d)$); 
				\node at ($(d)-(0,2*#4)$) {{\color{black} $#7$}};		
				}	

\newcommand{\coordSys}[9]{
			\coordinate (center2) at (#1,#2); 
            \coordinate (x1) at ($ (center2) + #3*(#4,#5) $); 
            \coordinate (x2) at ($ (center2) - .3*#3*(#4,#5) $); 
            \coordinate (y1) at ($ (center2) + #3*(#6,#7) $);
            \coordinate (y2) at ($ (center2) - .3*#3*(#6,#7) $);
            \coordinate (z1) at ($ (center2) + #3*(#8,#9) $); 
            \coordinate (z2) at ($ (center2) - .3*#3*(#8,#9) $); 
            \draw[->,>=stealth,black][line width=0.4mm] (x2) -- (x1);
			\node at ($(x1)+(0.2,0)$){{\color{black} \large $x$}};
            \draw[->,>=stealth,black][line width=0.4mm] (y2) -- (y1);
            \node at ($(y1)+(0,0.2)$){{\color{black} \large $y$}};
            \draw[->,>=stealth,black][line width=0.4mm] (z2) -- (z1);
			\node at ($(z1)+(0,0.2)$){{\color{black} \large $z$}};
    	}
    	
\newcommand{\coordSysTwoDim}[7]{
			\coordinate (center2) at (#1,#2); 
            \coordinate (x1) at ($ (center2) + #3*(#4,#5) $); 
            \coordinate (x2) at ($ (center2) - .3*#3*(#4,#5) $); 
            \coordinate (y1) at ($ (center2) + #3*(#6,#7) $);
            \coordinate (y2) at ($ (center2) - .3*#3*(#6,#7) $);
            \draw[->,>=stealth,black][line width=0.4mm] (x2) -- (x1);
			\node at ($(x1)+(0.2,0)$){{\color{black} \large $x$}};
            \draw[->,>=stealth,black][line width=0.4mm] (y2) -- (y1);
            \node at ($(y1)+(0,0.2)$){{\color{black} \large $y$}};
            }

\newcommand{\legRectTxt}[3]{
		\begin{tikzpicture}
			     \coordinate (center1) at (#1,#2); 
                \coordinate (b) at ($ (center1) - (.075,0) $);
                \coordinate (d) at ($ (center1) + (.075,0) $);
				 \draw[black,fill=#3] ($(b)-3*(0.1,0.02)$) rectangle ($ (d) + 3*(0.1,0.02) $); 
		\end{tikzpicture}		
}
\newcommand*{\Labelxy}[4]{\put(#1,#2) {\setlength{\fboxsep}{0pt}{\strut\textcolor{black}{\begin{turn}{#3}{#4}\end{turn}}}}}
\newcommand*{\LabelFig}[3]{\put(#1,#2) {\setlength{\fboxsep}{0pt}\colorbox{black}{\textcolor{white}{#3}}} }
\newcommand*{\LabelFigg}[3]{\put(#1,#2) {\setlength{\fboxsep}{2pt}\colorbox{white}{\textcolor{white}{#3}}} }
\definecolor{darkolivegreen}{rgb}{0.33, 0.42, 0.18}
\definecolor{darkspringgreen}{rgb}{0.09, 0.45, 0.27}
\definecolor{darkslategray}{rgb}{0.18, 0.31, 0.31}
\addeditor{KK}
\addeditor{SP}

\begin{document}

\title{Shear banding instability in \remove{high entropy} multi-component metallic glasses: Interplay of composition and short-range order}

\author{Kamran Karimi$^1$}
\email{Corresponding author: kamran.karimi@ncbj.gov.pl}
\author{Amin Esfandiarpour$^1$}%
\author{Ren\'e Alvarez-Donado$^1$}
\author{Mikko J. Alava$^{1,2}$}
\author{Stefanos Papanikolaou$^1$}
\affiliation{%
 $^1$ NOMATEN Centre of Excellence, National Centre for Nuclear Research, ul. A. Sołtana 7, 05-400 Swierk/Otwock, Poland\\
 $^{2}$ Aalto University, Department of Applied Physics, PO Box 11000, 00076 Aalto, Espoo, Finland
}%

\begin{abstract}
The shear-banding instability in quasi-statically driven bulk metallic glasses emerges from collective dynamics, mediated by shear transformation zones and associated non-local elastic interactions. It is also phenomenologically known that sharp structural features of shear bands are typically correlated to the sharpness of the plastic yielding transition, being predominant in commonly studied alloys composed of multiple different elements, that have very different atomic radii. However, in the opposite limit \remove{of high-entropy multicomponent alloys,} where elements' radii are relatively similar, plastic yielding of bulk metallic glasses is highly dependent on compositional and ordering features. 
In particular, a known mechanism at play involves the formation of short-range order dominated by icosahedra-based clusters.
Here, we report on atomistic simulations of multi-component metallic glasses with different chemical compositions showing that the degree of strain localization is largely controlled by the interplay between composition-driven icosahedra-ordering and collectively-driven shear transformation zones. By altering compositions, strain localization ranges from diffuse homogenized patterns to singular crack-like features. 
We quantify the dynamical yielding transition by measuring the atoms' susceptibility to plastic rearrangements, strongly correlated to the local atomic structure.
We find that the abundance of short-range ordering of icosahedra within rearranging zones increases glassy materials' capacity to delocalize strain. The kind of plastic yielding can be often qualitatively inferred by the commonly used compositional descriptor that characterizes element associations, the misfit parameter $\delta_a$, and also by uncommon ones, such as shear-band width and shear-band dynamics' correlation parameters.
\end{abstract}

\maketitle
\newpage
\section{Introduction}

Failure in sheared metallic glasses (below the glass transition temperature $T_g$) generically occurs via localization of intense, irrecoverable deformation beyond its elastic limit, commonly termed as shear bands \cite{maass2015shear}. 
These system-spanning strain features have typically narrow linear (in two dimensions) or planar (in three dimensions) topology along which most of plastic deformation accumulates, while the bulk of the material remains essentially elastically deformed. The underlying microstructural, dynamical mechanism that leads to shear-band nucleation and propagation in glassy metals has been known to follow a generally accepted picture in that the bulk plastic response emerges from \emph{collective} dynamics characterized by Shear Transformation Zones (STZs) \cite{argon1979plastic,falk1998dynamics, falk2011deformation, karimi2018correlation}. 
Nevertheless, in realistic systems, such as metal alloys, this picture has been thought to be valid in the so-called large-atomic disparity limit, characterized by the misfit coefficient $\delta_a$~\cite{senkov2001effect}, with $\delta_a>0.06$ being the \emph{good}-glass-forming limit where the atomic radii of compositional elements are considered to be very different \cite{li2019mechanical}. 
In this limit, local atomic displacements emerge spatially disordered, and under shear, they resemble the displacements seen in a wider range of soft glassy solids (i.e. jammed granular media, colloidal glasses, glassy polymers, or foams) within the context of plastic yielding transitions~\cite{ozawa2018random,denisov2017universal,karimi2017inertia,karimi2019plastic}. 
Nonetheless, there is a way to generate a glassy environment in the opposite limit of $\delta_a \rightarrow 0$ but maintaining a larger number of elements, as in \add{medium/}high entropy metallic glasses (see \cite{ding2013high} and references therein). In this multicomponent limit, the disorder in atomic displacements emerges as well, but the origin is in the ``color" of the atoms, \add{meaning the complexity of enthalpic interactions}, as opposed to their size.
In this limit, the behavior under shear has been a challenging concept that remains unexplored. In this paper, we focus on several multicomponent alloys that have been considered in the literature as medium- and high-entropy alloys~\cite{li2018microstructures,li2019mechanical,frydrych2021materials}, and we compare for their mechanical properties in their corresponding glassy state, which we achieve by fast cooling~\cite{lu2017cooling}. In an effort to understand which element/compositional and microstructural features signify the quality of the emergent glass under fast cooling rates, we investigate the correlation between shear-band microstructures with the average stress-strain behavior, and then associate them to the compositional structure. We find that plastic rearrangements strongly correlate to the local atomic structure and the abundance of short-range ordering of icosahedra within rearranging zones increases glassy materials' capacity to delocalize strain. The misfit parameter $\delta_a$ is correlated to the overall compositional trend, but also microstructural parameters, such as the emergent shear-band width and shear-band dynamics' correlation, \add{as well as a bulk order parameter tied to the sharpness of plastic yielding transition}.

It is currently clear that almost every basic aspect of bulk metallic glasses deformation and failure (e.g. plasticity, shear localization, preparation dependence, strain rate sensitivity, and thermal effects) have been thoroughly understood based on the STZ mechanism \cite{schuh2007mechanical}.
Moreover, the general picture concerning the role of STZs in deforming glassy alloys has been further validated universally across. 
This is remarkable given the inherent disparity of scales, microscopic interactions, and relaxation mechanisms involved.
Due to structural and/or mechanical) heterogeneity, the coalescence of atomic-scale STZs typically occurs in a highly intermittent and scattered manner both in time and space.
On approach to failure, STZ clusters restructure themselves over larger scales leading to the formation of co-existing macroscopic bands with a significant contribution to ductility and plastic flow \cite{wang2018spatial}.
On the other hand, certain (aged) glasses that lack this heterogeneity element \cite{fan2014evolution} (or associated lengths don't exceed interatomic scales) tend to localize strain within a single dominant band before the shear banding instablility results in a catastrophic brittle-type failure \cite{ma2016tailoring}.
Several studies highlighted the role of microstructural inhomogeneities in controlling the observed brittle-to-ductile transition.
In \cite{wang2008micromechanisms,das2005work,lee2006crystallization}, heterogeneities were mainly attributed to shear-induced crystallization introducing local hardening effects that hinder a dominant band propagation and, in turn, favor nucleation of multiple spontaneous shearing bands.
Similar properties were ascribed to quasi-crystal-like phases with short range order (SRO) featuring solid-like properties \cite{ding2014full,ding2012correlating,ding2014soft,ma2016tailoring} and, therefore, a strong tendency to tune the extent of strain localization \cite{shi2005strain,shi2006atomic,shi2007stress}.
SROs are indeed mediated by the formation of ordered icosahedral clusters as the most (energetically) favored atomic configuration in a quenched metallic glass which are also greatly responsible for a number of glassy properties such as the dynamical slow-down in the super-cooled regime, aging, and dynamic heterogeneity \cite{hufnagel2016deformation}.
We note that SROs serve as ``infertile" sites for the nucleation of STZs in that the latter are typically loosely-packed soft disordered arrangements that weaken local strength and enhance shear banding instabilities.

\begin{figure}[b]
    \centering
    \begin{overpic}[width=0.45\textwidth]{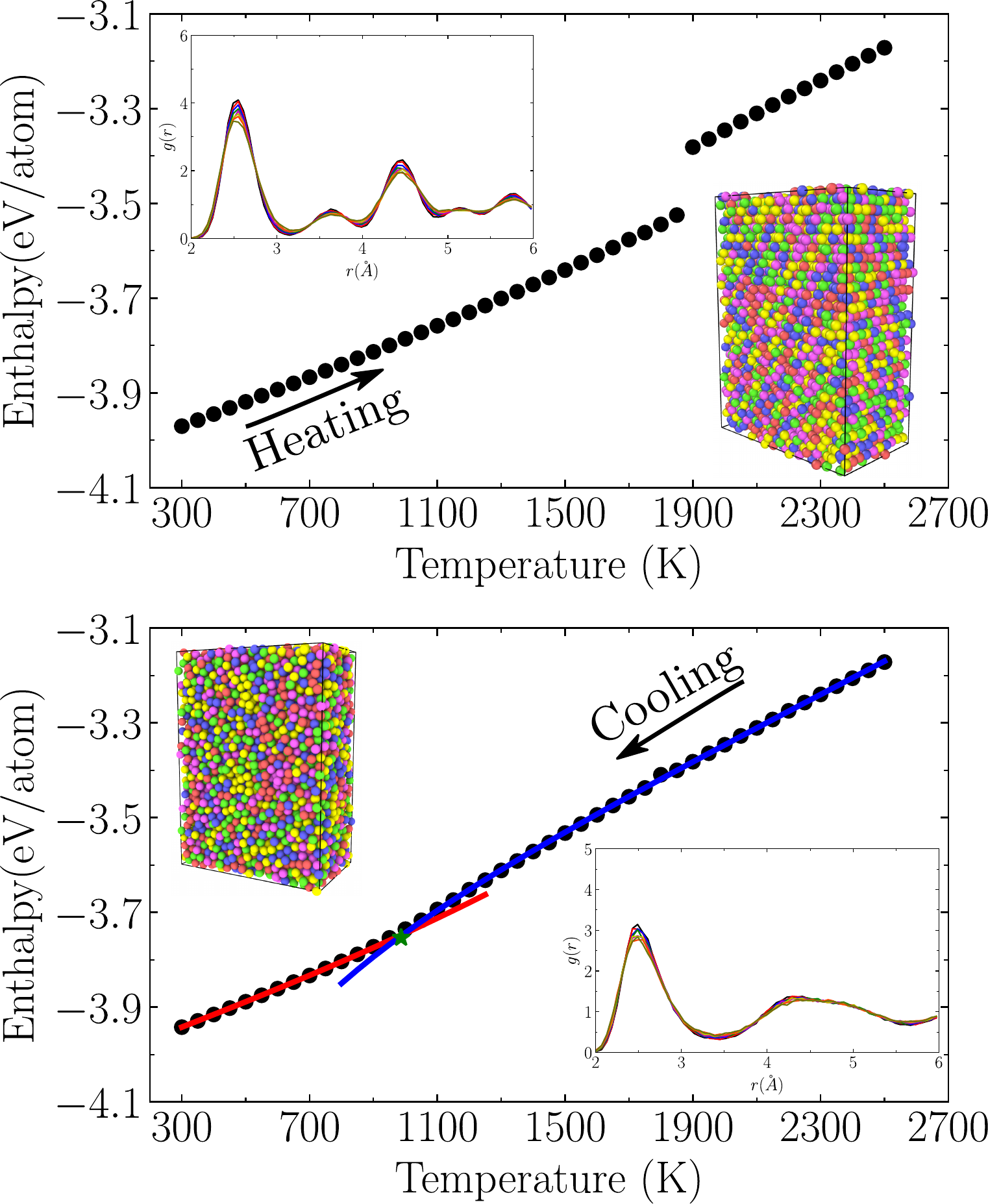}
        \LabelFig{13}{61}{$(a)$}
        \LabelFig{13}{10}{$(b)$}
	\end{overpic}
    \caption{Evolution of the per-atom enthalpy with temperature for CoNiCrFeMn upon \textbf{a}) melting \textbf{b}) quenching at $\dot T=0.1$~K/ps.  The insets show snapshots of atoms and associated radial distribution functions $g(r)$ at \textbf{a}) $T<T_m$ \textbf{b}) $T<T_g$. Here $T_m\simeq 1900$~K denotes the melting temperature corresponding to the enthalpy jump in \textbf{a}). The intersection of the nonlinear fits within the solid and liquid phases \cite{alcock1993materials} yields an estimate for the glass transition temperature $T_g\simeq 1000$~K in \textbf{b}).}
    \label{fig:thermo}
\end{figure}

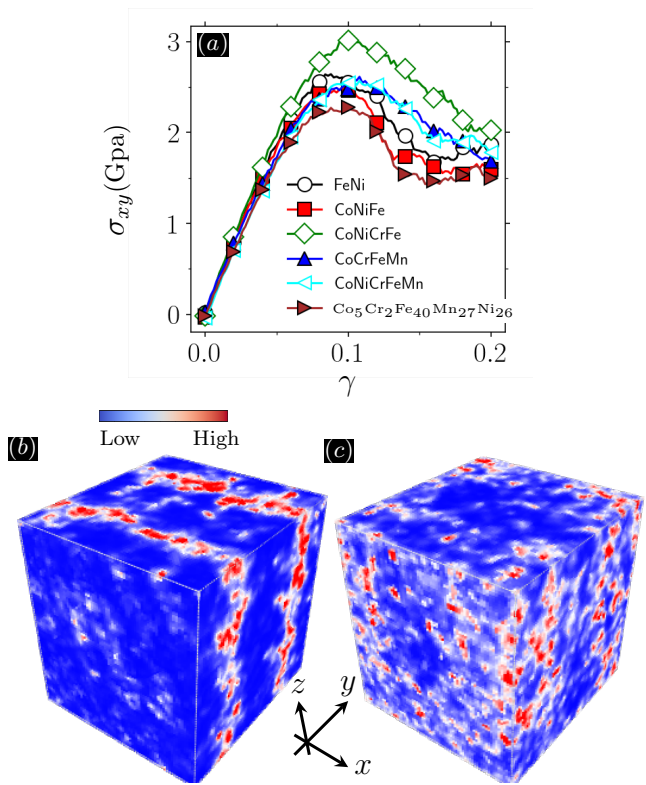
\begin{figure}
    \centering
    \begin{overpic}[width=0.3\textwidth]{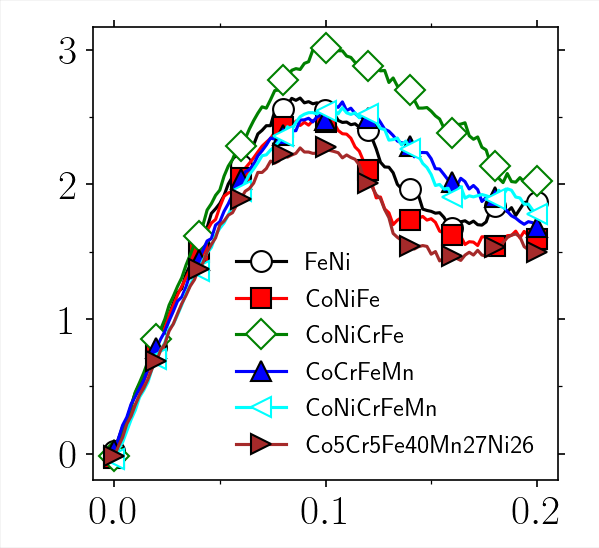}
        \LabelFig{17}{81}{$(a)$}
        \Labelxy{52}{-3}{0}{\large $\gamma$}
        \Labelxy{-6}{35}{90}{{\large $\sigma_{xy}$\normalsize (Gpa)}}
        \LabelFigg{51}{16}{\tiny Co$_5$Cr$_2$Fe$_{40}$Mn$_{27}$Ni$_{26}$}
        \Labelxy{50.5}{16}{0}{\tiny Co$_5$Cr$_2$Fe$_{40}$Mn$_{27}$Ni$_{26}$}
 \end{overpic}

    \vspace{28pt}
    \begin{overpic}[width=0.23\textwidth]{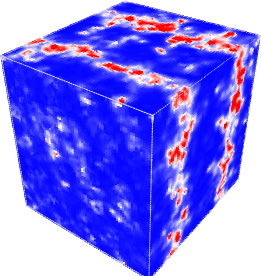}
        \LabelFig{-3}{100}{$(b)$}
        \put(25,110){\includegraphics[height=0.15cm,width=1.7cm]{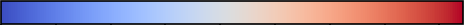}}  
        \put(25,103) {\footnotesize Low~~~~~~~High}
       \begin{tikzpicture}
            \coordinate (a) at (0,0); 
             \node[white] at (a) {\tiny.};
             \coordSys{3.7}{0}{.55}{1}{-0.6}{1}{1}{-0.2}{1} 
        \end{tikzpicture}	 
    \end{overpic}
    \begin{overpic}[width=0.23\textwidth]{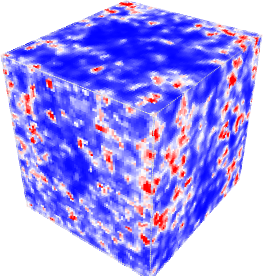}
        \LabelFig{-3}{100}{$(c)$}
       \begin{tikzpicture}
             \coordinate (a) at (0,0); 
             \node[white] at (a) {\tiny.};                 
    	\end{tikzpicture}	 
	\end{overpic}
    \caption{\textbf{a}) Macroscopic stress $\sigma_{xy}$ plotted against applied (shear) strain $\gamma$ corresponding to several metallic glasses. Local (interpolated) squared nonaffine displacements maps associated with the \textbf{b}) Co$_5$Cr$_2$Fe$_{40}$Mn$_{27}$Ni$_{26}$ and \textbf{c}) CoNiCrFe glasses at $\gamma=0.2$. Here $x$, $y$, and $z$ denote flow, gradient, and vorticity directions, respectively.}
    \label{fig:loadCurve}
\end{figure}


The above big picture, however, fails to provide specific  details on the precise nature of SRO-induced inhomogeneities and clear description as to how structural heterogeneities evolve at atomic scales and control the shear banding dynamics on macroscopic levels.
In this framework, our study aims to establish robust microstructural origins associated with varying degrees of shear localization in driven glassy alloys.
Our approach will be based on two (qualitative) themes relevant to the nature of the plastic yielding transition and \emph{quality} of metallic glasses: \emph{\romannum{1}}) a ``good" glass with a high ability to form localized deformation patterns \emph{\romannum{2}}) a ``bad" glass with a tendency to delocalize strain.   
\add{This might differ from the conventional notion based on the glass forming ability \cite{inoue2000stabilization} in that good glass formers (typically with notable $\delta_a$) do not necessarily deform well, according to our classification scheme.}
Our focus will be on certain types of local order which are purely structural in nature and strongly correlate with \emph{\romannum{1}}) dynamics of individual atoms \emph{\romannum{2}}) bulk stress response in good and bad glasses.   
The structural metrics used in this study include the density of local icosahedral clusters as well as the radial distribution function of individual atoms. 
The dynamical property is based on nonaffine displacements which should entail the atoms propensity to undergo localized rearrangements.
We also measure the spontaneous rate of stress release as a relevant mechanical property that helps discern a good glass from a bad one.  
Using a numerical framework, our work addresses the intimate structure-dynamics-property relationship by probing the local atoms environment, morphology of nucleated shear bands, and macroscopic stress relaxation and seek for potential correlations with the composition-based misfit parameter in sheared metallic glasses. 
The outcomes of our correlation analysis should describe the nature of the plastic yielding transition based upon relevant elemental/microstructural indicators with important implications in terms of materials tailoring to achieve desired mechanical properties.

The organization of the paper is as follows.
In Sec. \ref{sec:SimulationsAndProtocols}, the numerical setup, glass preparation, driving protocol, and relevant simulation details are discussed.
Section~\ref{sec:DiffuseVersusLocalized} presents the mechanical response corresponding to several simulated metallic glasses that, depending on chemical compositions, may nucleate diffuse or localized strain patterns.
In Sec. \ref{sec:StruDynProp}, we probe the structure-dynamics-property relationship through several structural and dynamical metrics. 
Section~\ref{sec:discussions} and \ref{sec:conclusions} present discussions and conclusions, respectively.

\section{Simulations and Protocols}\label{sec:SimulationsAndProtocols}
Molecular dynamics simulations were carried out in LAMMPS \cite{plimpton1995fast} by implementing atomistic systems of size $N=50,688$ within periodic cubic boxes with $80\le L\le 90$~\r{A} in a simple-shear loading geometry.
The model metallic glasses included in this study are FeNi, CoNiFe, CoNiCrFe, CoCrFeMn, CoNiCrFeMn, and Co$_5$Cr$_2$Fe$_{40}$Mn$_{27}$Ni$_{26}$.
The interatomic forces were derived from the modified embedded atom method (MEAM) potential with the input parameters associated with the principle elements obtained from \cite{choi2018understanding}.
The glassy samples were prepared based on the \remove{standard} melt-and-quench protocol illustrated in Fig.~\ref{fig:thermo} plotting the enthalpy of CoNiCrFeMn as a function of temperature.
In this context, a \remove{(nearly)} crystalline structure was initially heated up to a high temperature ($3000$~K) \add{above the melting temperature $T>T_m$}, as in Fig.~\ref{fig:thermo}(a), \remove{far below $T_g$} before it was cooled down to the room temperature ($300$~K) below $T_g$ at the quench rate of $\dot{T}= 0.1$~K/ps, as in Fig.~\ref{fig:thermo}(b).
We allowed for further relaxation upon quenching for order $10$ ps to ensure as-quenched glasses are fully equilibrated prior to shearing.
The ambient pressure $P_0$ was set to zero at all times.
We also set the discretization time to $\Delta t=0.001$ ps.
The NPT ensembles were implemented via a Nose-Hoover thermostat and barostat with relaxation time scales $\tau_d^\text{therm}=0.2$~ps and $\tau_d^\text{bar}=2.5$~ps. 
Prior to shearing, the bulk stress tensor $\sigma_{\alpha\beta}\simeq-P_0~ \delta_{\alpha\beta}$ has a nearly hydrostatic form.
A strain-controlled condition was applied by deforming the periodic box on the $xy$ plane \add{implementing Lees-Edwards boundary conditions \cite{lees1972computer}} at a fixed volume and temperature (canonical NVT ensembles) and a constant shearing rate $\dot\gamma=10^{-4}$ ps$^{-1}$ up to a $20\%$ strain.

\section{Deformation Patterns: Diffuse Vs. Localized}\label{sec:DiffuseVersusLocalized}
We performed a series of shear tests on the quenched samples with the resulting load curves $\sigma_{xy}$ against applied shear strain $\gamma$ reported in Fig.~\ref{fig:loadCurve}(a).
In all the simulated glasses, the stress rises monotonically toward a peak stress followed by a sheer reduction in strength as the loading continues.
We note that the Co$_5$Cr$_2$Fe$_{40}$Mn$_{27}$Ni$_{26}$ metallic glass exhibits a very pronounced and abrupt stress drop that is typically accompanied by sharp strain features as illustrated in Fig.~\ref{fig:loadCurve}(b).
The blue and red colors in the deformation maps indicate regions with low and high squared nonaffine displacements (see Sec.~\ref{sec:StruDynProp}).
The FeNi alloy almost shows identical stress features with an exception that the post-yielding behavior occurs in a smoother manner.
These metals should be referred to as good glasses with a high susceptibility to localize deformation.
The other compositions, the so-called bad glasses, seem to indicate a shallower stress decay that is associated with rather diffuse deformation patterns across the medium, as in Fig.~\ref{fig:loadCurve}(c).

Intuitively, the blue matrix in the deformation maps serve as fertile spots for the nucleation of SROs whereas the red (hot) regions should be structurally rich in terms of STZ clusters.
Yet, there is an evolving fraction of icosahedral arrangements that tend to co-exist with STZs within the hot spots.
We argue that the observed co-existence indeed contributes to structural heterogeneities associated to the sheared glass and, effectively, controls its quality (i.e goodness versus badness) upon failure.

\begin{figure}[b]
    \centering
    \begin{overpic}[width=0.3\textwidth]{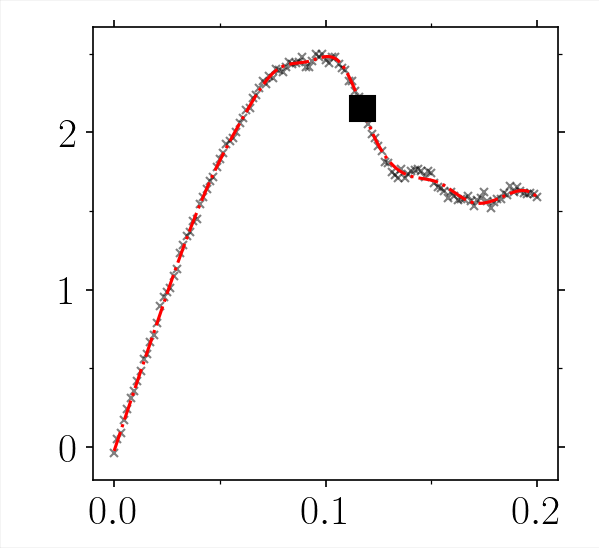}
        \Labelxy{50}{-3}{0}{\large $\gamma$}
        \Labelxy{-6}{35}{90}{\large $\sigma_{xy}$\normalsize (Gpa)}
       \begin{tikzpicture}
            \coordinate (a) at (0,0); 
            \node[white] at (a) {\tiny.};               %
                \drawSlope{3.0}{3.4}{0.4}{20}{black}{\hspace{-8pt}\large h_\text{min}}{\vspace{2pt}1}
			\coordinate (center2) at (3.1,3.8); 
            \coordinate (b) at ($ (center2) - .3*(1,-2.3) $); 
            \coordinate (c) at ($ (center2) + .3*(1,-2.3) $);
            \draw[black][line width=0.4mm] (c) -- (b); 
            \coordinate (d) at ($ (center2) - .5*(0,1) $); 
		    \end{tikzpicture}
		    
	\end{overpic}
    \caption{Illustration of the softening modulus $h_\text{min}$. The (black) crosses indicate the stress data $\sigma_{xy}$ as a function of the applied (shear) strain $\gamma$. The dashdotted (red) curve denotes a nonlinear fit based on (cubic) smoothing splines.  The square symbol indicates the strain corresponding to the minimal slope past the peak as illustrated by the straight line.}
    \label{fig:hminIllustration}
\end{figure}
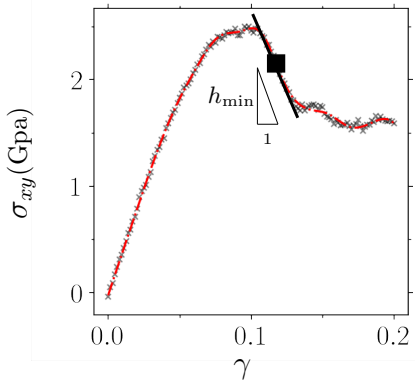

\section{Structure-dynamics-property relation}\label{sec:StruDynProp}
We now turn to relevant structural and dynamical metrics and search for possible connections with glass properties.
As a starting point, dynamics of individual atoms is quantified by $D^2_\text{min}$ as a measure of deviations from affine trajectories of atoms imposed by an external (homogeneous) shear \cite{falk1998dynamics}.
We fit a linear model based on the reference positions $\vec{\mathring{r}}$ at zero strain, i.e $\vec{\mathcal{R}}(\vec{\mathring{r}})={\bf{F}} \vec{\mathring{r}}+\vec{F}_0$, to the current atomic positions $\vec{r}$ associated with $N_b$ atoms within a given volume \footnote{The $D^2_\text{min}$ analysis involves partitioning of the entire box into sub-volumes using a cubic grid of size $r_c=3.0$~\r{A} which is close to twice the radius associated with the first peak of $g(r)$.} by minimizing the squared loss function $\mathcal{L}({\bf{F}},
\vec{F}_0)=\sum_{i=1}^{N_b}|\vec{r}_i-\vec{\mathcal{R}}(\vec{\mathring{r}}_i)|^2$ to obtain $\{\hat{\bf{F}},\hat{\vec{F}}_0\}=\text{argmin}~\mathcal{L}({\bf{F}},
\vec{F}_0)$.
Here the second-ranked deformation tensor $\bf{F}$ and the (rigid) translation vector $\vec{F}_0$ correspond to an affine transformation.
The squared nonaffine displacements for atom $i$ is defined as $D^2_{i}=|\vec{r}_i-\hat{\vec{r}}_i|^2$ with $\hat{\vec{r}}_i$ being the predicted position $\hat{\vec{\mathcal{R}}}(\vec{\mathring{r}})={\hat{\bf{F}}} \vec{\mathring{r}}+\hat{\vec{F}}_0$ due to affine deformations.

Next, we performed a Voronoi analysis using OVITO \cite{stukowski2009visualization} to locate atoms with an icosaheral symmetry---that is, Voronoi cells with exactly 12 polyhedral faces with 5 edges each.
To obtain the associated (number) density, $\rho_i^\text{ico}=1/V_i^\text{ico}$, we repeated the Voronoi analysis by including \emph{exclusively} atom $i$'s with icosahedral symmetries within the periodic box (and excluding the other atoms).
This gives another set of Voronoi cells (not to be confused with the original Voronoi set) with volume $V_i^\text{ico}$ of atom index $i$.
We further employed the other structure characterization methods based on the radial distribution function $g(r)$ and strain localization width $w_\text{sb}$. 
Finally, the glass quality, as a mechanical property, was gauged based on the softening modulus 
\begin{equation}
h_\text{min} = -\partial_\gamma \sigma_{xy}\Big|_{\rm min}
\end{equation}
denoting the rate of stress decay associated with the maximum instability state upon failure (see Fig.~\ref{fig:hminIllustration}).
In the following subsections, we probe inter-correlations between the above metrics and outline commonalities and differences based on our ``good-versus-bad" theme.

\begin{figure}
    \raggedright
    \begin{overpic}[width=0.27\textwidth]{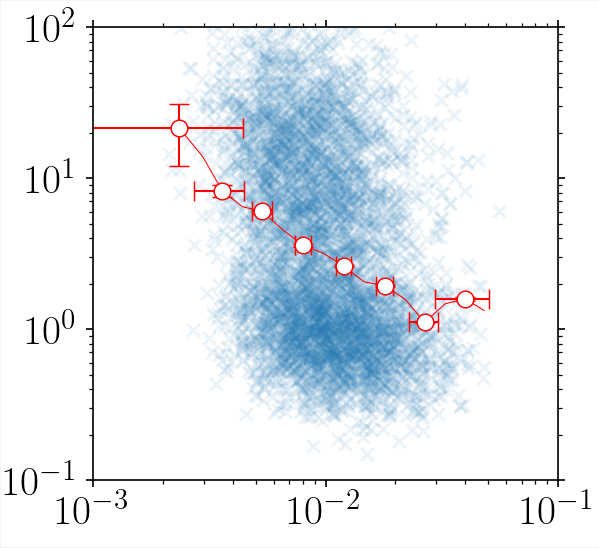}
        \Labelxy{-8}{35}{90}{$D^2_\text{min}$ (\r{A}$^2$)}
        \Labelxy{17}{15}{0}{$c_{XY}=-0.4$}
         \LabelFig{17}{80}{$(a)$ \tiny Co$_5$Cr$_2$Fe$_{40}$Mn$_{27}$Ni$_{26}$}
         \put(116,10){\includegraphics[height=0.15cm,width=1.7cm]{Figs/colorBar_new.png}}
         \put(116,4) {$\footnotesize \text{Low~~~~~~~High}$}
         \put(106,19){\includegraphics[width=0.15\textwidth]{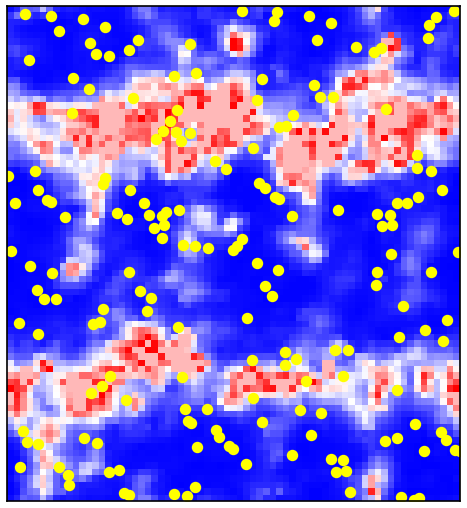}}
    \end{overpic}

     \begin{overpic}[width=0.27\textwidth]{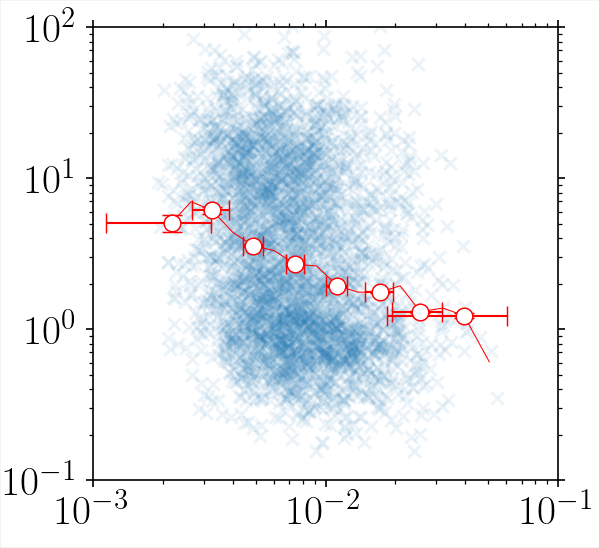}
        \Labelxy{38}{-3}{0}{$\rho_\text{ico}$(\r{A}$^{-3}$)}
        \Labelxy{-8}{35}{90}{$D^2_\text{min}$ (\r{A}$^2$)}
        \Labelxy{17}{15}{0}{$c_{XY}=-0.2$}
        \put(106,19){\includegraphics[width=0.15\textwidth]{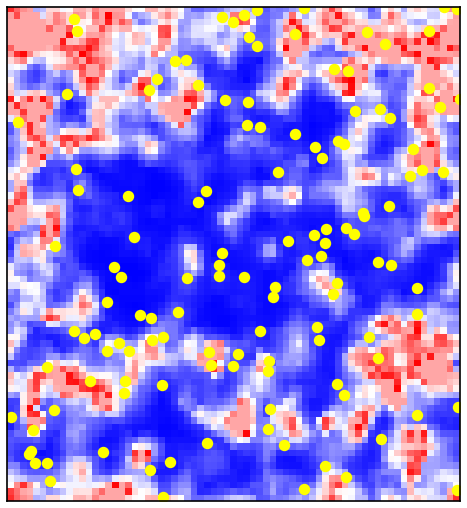}}
        \LabelFig{17}{80}{$(b)$ \tiny CoNiCrFe}
        \begin{tikzpicture}
             \coordinate (a) at (0,0); 
             \node[white] at (a) {\tiny.};                 
    		 \scaleBarr{6.2}{0.0}{0.35}{0.17}{$\footnotesize 0$}{$\footnotesize 20$}{$\footnotesize 40$}{\r{A}}
             \coordSysTwoDim{4.85}{0}{.55}{1}{0}{0}{1} 
        \end{tikzpicture}
    \end{overpic}
    \caption{Linking local icosahedral ordering with $D^2_\text{min}$ in the \textbf{a}) Co$_5$Cr$_2$Fe$_{40}$Mn$_{27}$Ni$_{26}$ and \textbf{b}) CoNiCrFe glasses at $\gamma=0.2$. The left panels display scatter plots of $D^2_\text{min}$ and icosahedra density $\rho_\text{ico}$ with the (red) symbols denoting binning averaged data (in log space). The corresponding $D^2_\text{min}$ maps are shown on the right with the blue and red colors indicating regions with low and high nonaffine displacements. The (yellow) dots denote atoms with full icosahedral order.}
    \label{fig:scatterD2minRho}
\end{figure}

\subsection{Structural \& Dynamical Descriptors}
The color maps in Fig.~\ref{fig:scatterD2minRho} overlay atoms with the full icosahedral order (yellow disks) on the two-dimensional (interpolated) $D^2_\text{min}$ field associated with the Co$_5$Cr$_2$Fe$_{40}$Mn$_{27}$Ni$_{26}$ (top) and CoNiCrFe (bottom) metallic glasses.
It is very clear on both maps that (red) rearranging regions significantly lack local structural ordering as opposed to the (blue) rigid matrix.
As a metric to quantify this observation, a cross correlation analysis was carried out between the squared nonaffine displacements of the yellow discs and the associated (number) density $\rho_\text{ico}$.
The scatter data of $D^2_\text{min}$ and $\rho_\text{ico}$ are shown in Fig.~\ref{fig:scatterD2minRho}(a, b) along with the binning-averaged data (in the logarithmic space).
The observed trend indicates significant (anti-)correlations between the logarithm of the two observables $X=\text{log}_{10}{D^2_\text{min}}$ and $Y=\text{log}_{10}\rho_\text{ico}$ with the (linear) correlation coefficient $c_{XY}\simeq -0.4, -0.2$ associated with Co$_5$Cr$_2$Fe$_{40}$Mn$_{27}$Ni$_{26}$ and CoNiCrFe, respectively. 
Here $c_{XY}=\langle~\hat{X}_{i}.\hat{Y}_{i}\rangle$ where $\langle.\rangle_i$ denotes averaging over the atom index $i$ and $\hat{X}$ indicates the fluctuating part (with the mean value subtracted) normalized by the standard deviation associated with each variable.

\begin{figure}[t]
    \begin{overpic}[width=0.27\textwidth]{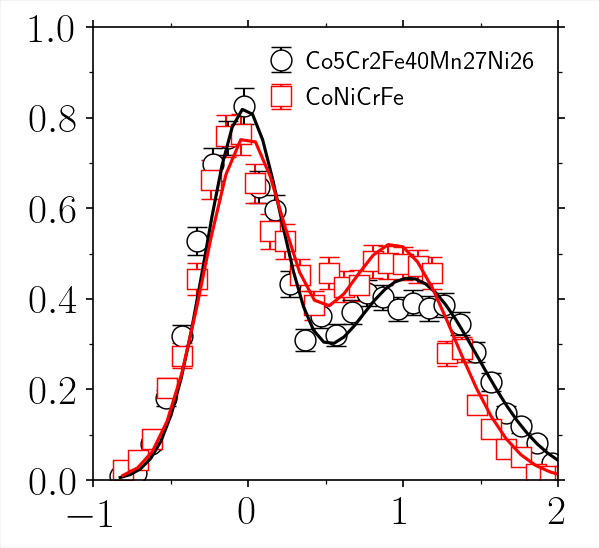}
        \Labelxy{-8}{27}{90}{$P(\text{Log}_{10} D^2_\text{min})$}
        \Labelxy{38}{-3}{0}{$\text{Log}_{10} D^2_\text{min}$}
        \begin{tikzpicture}
            \coordinate (a) at (0,0); 
            \node[white] at (a) {\tiny.};               %
			\coordinate (center2) at (2.55,1.0); 
            \coordinate (b) at ($ (center2) + .1*(1,0) $); 
            \coordinate (c) at ($ (center2) - .1*(1,0) $);
            \draw[red][line width=0.2mm] (c) -- (b); 
            \coordinate (d) at ($ (center2) - .5*(0,1) $); 
            \draw[->,>=stealth,red][line width=0.2mm] (center2) -- (d); 
		    \node at ($(center2)+(0.0,0.2)$) {{\color{red}$\scriptstyle \text{Log}_{10} D^2_\text{th}$}}; %
		    \end{tikzpicture}
    \end{overpic}
    \caption{$D^2_\text{min}$ probability distribution function of atoms with the full icosahedral order corresponding to the Co$_5$Cr$_2$Fe$_{40}$Mn$_{27}$Ni$_{26}$ and CoNiCrFe glasses at $\gamma=0.2$. The solid curves indicate results of the Gaussian mixture model. $D^2_\text{th}$ indicated by the arrow separates the two populations. The error bars denote standard errors.}
    \label{fig:pdfCondD2min}
\end{figure}
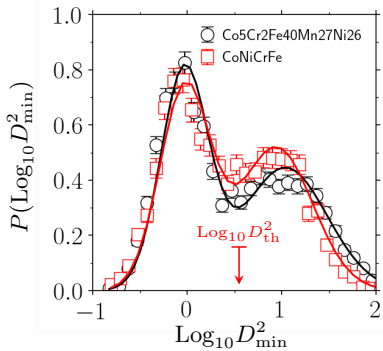

Stronger anti-correlations associated with the ``good" glass imply an infrequent occurrence of the structural ordering within the deforming bands potentially indicative of a strain-softening mechanism. 
The ``bad" glass, by contrast, favors a spontaneous formation of (relatively) denser icosahedral clusters inside shear zones with partial strengthening effects that may account for the shallow stress decay associated with its bulk response.
Variations in the relative fraction of ordered clusters is also manifested in the corresponding $D^2_\text{min}$ probablity distribution function.
As shown in Fig.~\ref{fig:pdfCondD2min}, both glasses exhibit a clear bimodal behavior with the first higher and second lower peaks denoting the population of icosahedral clusters forming outside and within  shear zones, respectively. 
Results of the Gaussian mixture model associated with the (bad) CoNiCrFe glass (the red squares) indicates a relatively higher contribution of the ordered icosahedral phase to the plastic flow in comparison with the (good) Co$_5$Cr$_2$Fe$_{40}$Mn$_{27}$Ni$_{26}$ metallic glass (the black circles).
This observation is also in line with the outcomes of our correlation analysis.

\begin{figure}[b]
    \raggedright
    \begin{overpic}[width=0.27\textwidth]{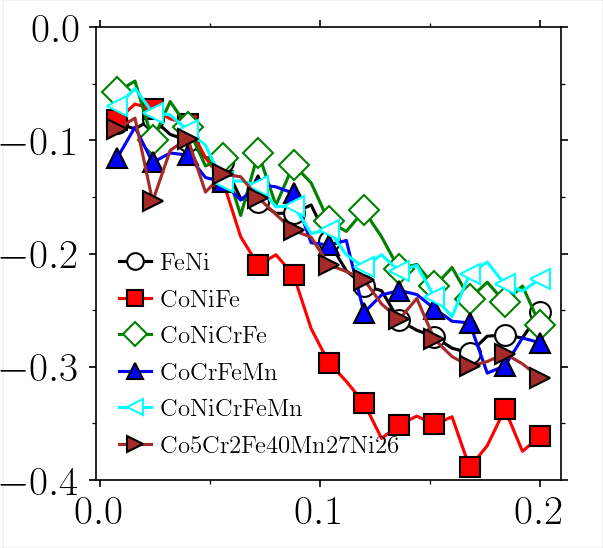}
        \Labelxy{50}{-3}{0}{\large $\gamma$}
        \Labelxy{-8}{42}{90}{\large$c_{XY}$}
         \LabelFig{17}{80}{$(a)$}
         \put(96,9){\includegraphics[width=0.22\textwidth]{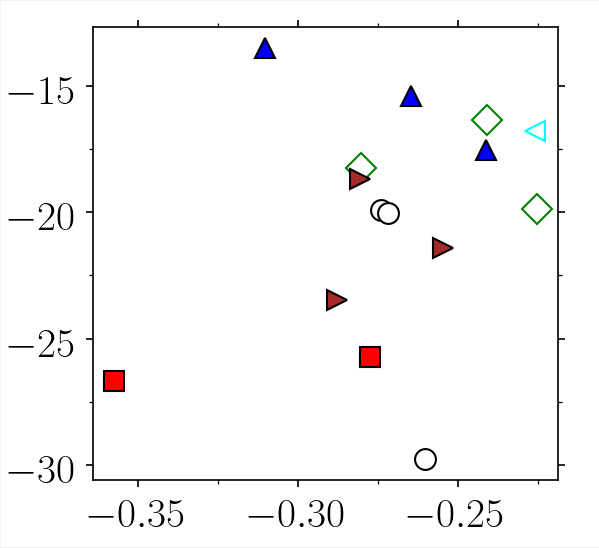}}
         \LabelFig{110}{74}{$(b)$}
        \Labelxy{175}{34}{90}{ $h_\text{min}$\footnotesize~(Gpa)}
        \Labelxy{132}{-3}{0}{\large$c^\infty_{XY}$}
    \end{overpic}
    \caption{\textbf{a}) Evolution of the correlation coefficient $c_{XY}$ with applied strain $\gamma$ in several metallic glasses. \textbf{b}) Scatter plot of the softening modulus $h_{min}$ and (terminal) correlation value $c^\infty_{XY}$ associated with different glasses. Each symbol on the right denotes a set of different realizations corresponding to a specific glass composition.}
    \label{fig:cxyGamma}
\end{figure}

Our analysis suggests that the interplay between the shear zones dynamics and nucleation of ordered icosahedral clusters might indeed control the extent of shear banding instability in metallic glasses.
In this framework, we probed the evolution of such correlations with the applied strain as well as their variation with glass compositions.
Figure~\ref{fig:cxyGamma}(a) illustrates that the degree of (anti-)correlations between $D^2_\text{min}$ and $\rho_\text{ico}$ grows monotonically with strain by the time it saturates at $\gamma\simeq0.2$ but also reveals meaningful variations with the glass quality.
As shown in Fig.~\ref{fig:cxyGamma}(b), the latter, quantified by $h_\text{min}$, strongly correlates with the ultimate correlation value where better glasses (with lower $h_\text{min}$) correspond, on average, to a lower $c^\infty_{XY}$.  

\begin{figure}[t]
    \centering
    \begin{overpic}[width=0.27\textwidth]{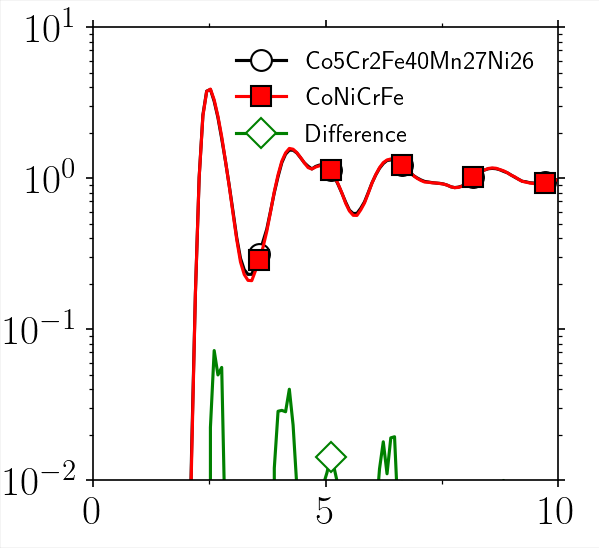}
        \Labelxy{50}{-3}{0}{\large $r$\normalsize (\r{A})}
        \Labelxy{-8}{42}{90}{\large $g(r)$\normalsize}

		\end{overpic}
    
    \caption{Pair distribution function $g(r)$ of the Co$_5$Cr$_2$Fe$_{40}$Mn$_{27}$Ni$_{26}$ and CoNiCrFe glasses at $\gamma=0.2$ conditioned based on $D^2_\text{min}>D^2_\text{th}$ (see Fig.~\ref{fig:pdfCondD2min}) associated with center atoms.}
    \label{fig:grD2min}
\end{figure}

Another structural metric we probed is the radial distribution function which should allow for discerning local structural differences leading to the observation of fundamentally different dynamics.
As for the bad glasses, we speculate that the favorable icosahedra formation within plastic zones may have a signature on local density fluctuations around highly rearranging atoms.
We make an appropriate conditioning on $g(r)$ by including center atoms $i=1...N_\alpha$ with $D^2_\text{min}>D^2_\text{th}$, i.e.
\begin{equation}
g(r)=\frac{1}{4\pi r^2\Delta rN_\alpha\bar{\rho}}\sum_{i=1}^{N_\alpha}\sum_{j=1}^{N} \delta(r-r_{ij}), 
\end{equation}
with the mean number density of the system $\bar{\rho}$, discretization distance $\Delta r$, Kronecker delta $\delta(.)$, and pair-wise distance $r_{ij}$ of atoms $i$ and $j$. 
Here the chosen threshold $D^2_\text{th}$ separates the two $D^2_\text{min}$ populations as in Fig.~\ref{fig:pdfCondD2min}.
Figure~\ref{fig:grD2min} plots the conditioned $g(r)$ associated with the Co$_5$Cr$_2$Fe$_{40}$Mn$_{27}$Ni$_{26}$ and CoNiCrFe glasses as well as the difference between that of the latter and the former.
We note that, despite significant differences in icosahedral ordering and dynamic properties, the two metals have almost identical pair distribution functions. 
Nevertheless, the first few characteristic peaks in $g(r)$ corresponding to CoNiCrFe are slightly higher than those of Co$_5$Cr$_2$Fe$_{40}$Mn$_{27}$Ni$_{26}$ (see the green diamonds). 
This may indicate a relatively stronger presence of short-range ordering within shear spots of the former glass which were shown to be richer (relative to the latter metal) in terms of the density of icosahedral clusters.

\subsection{Shear Band Characterization}

Naturally, topological features associated with the shear band nucleation in metallic glasses should have have a strong bearing on the dynamics of stress near the failure point.
Along these lines, we made measurements of shear band widths $w_{sb}$ as a useful metric to quantify the degree of strain localization and sought for potential correlations with the softening modulus $h_\text{min}$. 
In the context of an ideal plastic behavior (with $h_\text{min}\simeq 0$), flow is accompanied by a nucleation of multiple small-scale bands that tend to spread uniformly all across the system.
In this case, one may assume that the shear band width approaches the system size, i.e. $w_\text{sb}\simeq L$.
On the other hand, brittle fracture is typically associated with extremely localized and crack-like deformation features often leading to singular stress curve (with $h_\text{min} \rightarrow -\infty$ in the theoretical limit) \cite{ozawa2018random}.
The following aims to explore such an intimate relationship between $w_{sb}$ and $h_\text{min}$ in the context of metallic glasses.

\begin{figure}[b]
    \raggedright
    \begin{overpic}[width=0.27\textwidth]{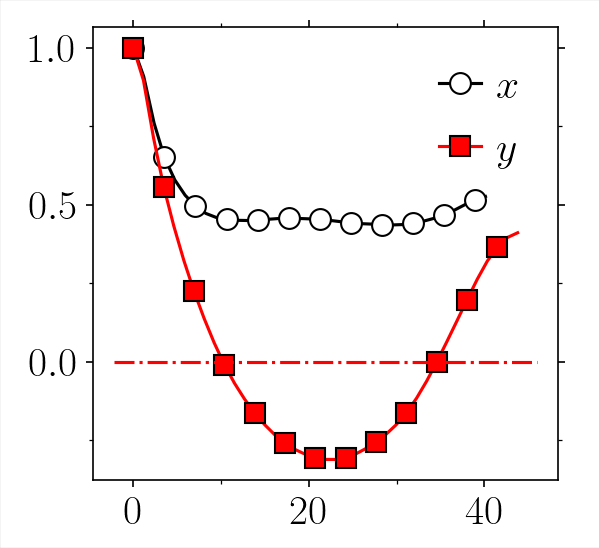}
        \LabelFig{-3}{90}{$(a)$ \tiny Co$_5$Cr$_2$Fe$_{40}$Mn$_{27}$Ni$_{26}$}
        \Labelxy{-8}{50}{90}{\large $c(r)$\normalsize}
         \put(100,12){\includegraphics[width=0.18\textwidth]{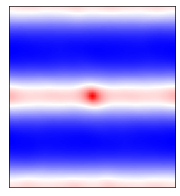}}
        \begin{tikzpicture}
             \coordinate (a) at (0,0); 
             \node[white] at (a) {\tiny.};                 
     		 \scaleBarr{6.0}{0.0}{0.35}{0.17}{$\footnotesize 0$}{$\footnotesize 20$}{$\footnotesize 40$}{\r{A}}
             \coordSysTwoDim{6.35}{1.95}{.55}{1}{0}{0}{1} 
        \end{tikzpicture}

    		 \end{overpic}
    \begin{overpic}[width=0.27\textwidth]{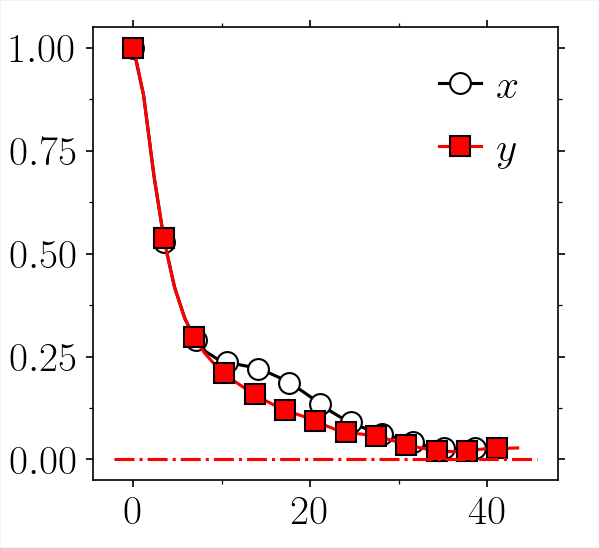}
        \LabelFig{-3}{90}{$(b)$ \tiny CoNiCrFe}
        \Labelxy{47}{-3}{0}{\large $r$\normalsize (\r{A})}
        \Labelxy{-8}{45}{90}{\large $c(r)$\normalsize}
         \put(100,12){\includegraphics[width=0.18\textwidth]{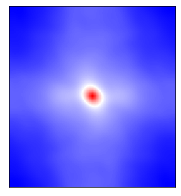}}
         \put(116,6){\includegraphics[height=0.15cm,width=1.7cm]{Figs/colorBar_new.png}}
         \put(116,0) {$\footnotesize \text{Low~~~~~~~High}$}
         \begin{tikzpicture}
             \coordinate (a) at (0,0); 
             \node[white] at (a) {\tiny.};                 
             \coordSysTwoDim{6.35}{2.25}{.55}{1}{0}{0}{1} 
        \end{tikzpicture}
    \end{overpic}
    
    \caption{$D^2_\text{min}$ correlations corresponding to the \textbf{a}) Co$_5$Cr$_2$Fe$_{40}$Mn$_{27}$Ni$_{26}$ and \textbf{b}) CoNiCrFe glasses at $\gamma=0.2$.  The correlation function $c(r)$ is plotted along the flow direction $x$ and gradient direction $y$. The correlation maps $c(x,y,z=0)$ are shown on the right. The flat (dashdotted) line indicate zero correlations. }
    \label{fig:strnCrltn}
\end{figure}

We performed an auto-correlation analysis on local $D^2_\text{min}$ fields to extract relevant micro-structural lengths.
We start by interpolating the squared nonaffine atomic displacements on a fine cubic grid  to obtain the auto-correlation function 
\begin{equation}
    c(\vec{r})=\langle~\hat D^2_\text{min}(\vec{r}^{\hspace{2pt}\prime}+\vec{r}).\hat D^2_\text{min}(\vec{r}^{\hspace{2pt}\prime})~\rangle,
\end{equation}
with $\hat{D^2_\text{min}} \doteq (D^2_\text{min}-\langle D^2_\text{min}\rangle)/\text{var}^{\frac{1}{2}}(D^2_\text{min})$.
Figure~\ref{fig:strnCrltn}(a, b) displays the correlation results corresponding to Co$_5$Cr$_2$Fe$_{40}$Mn$_{27}$Ni$_{26}$ and CoNiCrFe glasses at $\gamma=0.2$.
In Fig.~\ref{fig:strnCrltn}(a), $c(r)$ associated with the former alloy shows an early decay regime along the flow direction $x$ before entering a plateau regime at a bigger $r$ due to the shear band formation that spans across the system (see the map of Fig.~\ref{fig:scatterD2minRho}(a)).  

Along the gradient direction $y$, however, $c(r)$ exhibits a nonmonotonic behavior; it crosses zero at $r\simeq10$~\r{A} and reaches a minimum at a larger length implying the presence of strong anti-correlations.
The former length should give a reasonable width estimate for the nucleated bands, in visual agreements with the $D^2_\text{min}$ map of Fig.~\ref{fig:scatterD2minRho}(a).
Negative correlations may suggest a ``coexistence" between intensely deforming plastic zones and the surrounding solid-like matrix with negligible shear deformation.
The auto-correlation function becomes positive at $r>35$~\r{A} owing to the spontaneous formation of two shearing bands which are roughly apart by the indicated distance.
The CoNiCrFe glass, on the other hand, shows monotonic slowly-decaying correlations in both $x$ and $y$ directions hitting the noise floor at $r\simeq40$~\r{A} which is roughly half the physical system size.
The picture is fairly consistent with the spatial structure of plastic zones with a fairly homogeneous distribution of nonaffine deformations across the bulk, as in Fig.~\ref{fig:scatterD2minRho}(b). 

\begin{figure}[t]
    \centering
    \begin{overpic}[width=0.23\textwidth]{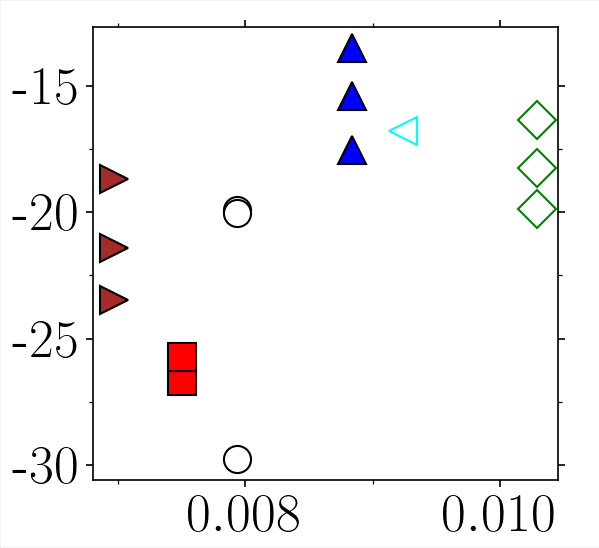}
        \LabelFig{17}{80}{$(a)$}
        \Labelxy{50}{-5}{0}{\large $\delta_a$}
        \Labelxy{-8}{26}{90}{\large $h_\text{min}$\normalsize (Gpa)}
    \end{overpic}
    \begin{overpic}[width=0.23\textwidth]{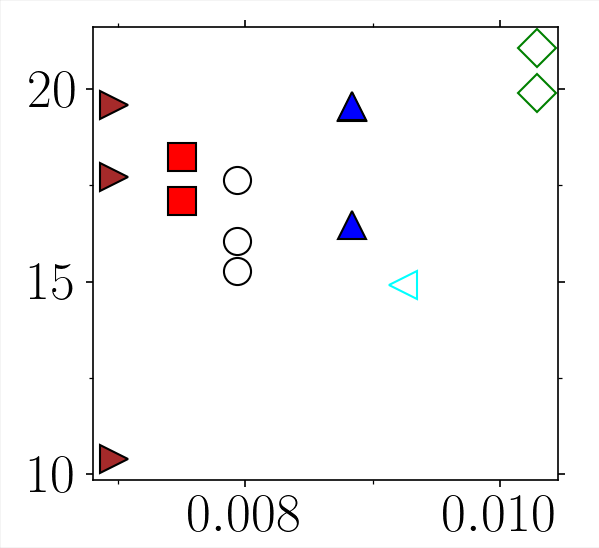}
        \LabelFig{17}{80}{$(b)$}
        \Labelxy{50}{-5}{0}{\large $\delta_a$}
        \Labelxy{97}{32}{90}{\large $w_\text{sb}$\normalsize (\r{A})}
    \end{overpic}
    \caption{Scatter plot of \textbf{a}) softening modulus $h_\text{min}$ and \textbf{b}) shear band width $w_\text{sb}$ plotted against the size mismatch parameter $\delta_a$ corresponding to the FeNi(\protect\circTxtFill{0}{0}{white}), CoNiFe(\protect\legSqTxt{0}{0}{red}), CoNiCrFe(\protect\legDiamondTxt{0}{0}{darkspringgreen}), CoCrFeMn(\protect\legTriangleTxt{0}{0}{blue}), CoNiCrFeMn(\protect\leftTriang{0}{0}{cyan}), Co$_5$Cr$_2$Fe$_{40}$Mn$_{27}$Ni$_{26}$ (\protect\rightTriang{0}{0}{brown}) metallic glasses.}
    \label{fig:SoftModuAndWidth}
\end{figure}

Figure~\ref{fig:SoftModuAndWidth} plots the $w_\text{sb}$ and $h_\text{min}$ estimates for different glasses against the corresponding atomic size mismatch parameter $\delta_a = \sqrt{\text{var}(a_i)}/\langle a_i\rangle$.
The latter is defined as the normalized standard deviation associated with radius $a_i$ of atom $i=1...N$.
In the context of high-entropy alloys, $\delta_a$ is believed to strongly control the degree of local lattice distortions \cite{shang2021mechanical}.
We generalize this concept to glassy alloys where the size mismatch parameter may be thought as a structural proxy controlling the degree of quenched disorder.
Here $\delta_a$ varies between $1-2\%$.
The scatter plot of the softening modulus $h_\text{min}$ and size mismatch $\delta_a$ in Fig.~\ref{fig:SoftModuAndWidth}(a) indicates a positive correlation between the two observables although the CoCrFeMn samples do not seem to follow the overall trend. 
Likewise, the shear band width $w_\text{sb}$ in Fig.~\ref{fig:SoftModuAndWidth}(b) seem to be statistically related with $\delta_a$.
Our data (not shown) also suggest that the glass goodness, featured by $h_\text{min}$, correlates with the localization degree measured by $w_\text{sb}$.

\section{Discussions}\label{sec:discussions}
Our atomistic simulations of sheared metallic glasses with several principal components have led to the observation of different strain patterns, ranging from a homogeneous plastic-like flow to shear band dominated localized deformation, in agreement with previous numerical and experimental studies. 
We have found a direct relevance of underlying microstructure on the atomic-scale dynamics which strongly correlates with the observed deformation features as well as the bulk mechanical response upon failure.
Our Voronoi-based analysis demonstrates the prevalence of the short-range order based on an icosahedral symmetry attributed to the local inter-atomic topology.
Atoms with the observed symmetry are found to be less prone to undergo intense shear transformations.
This has been validated by observing pronounced anti-correlation features between (atomic-scale) nonaffine displacements and density of icosahedral clusters. 
Our correlation analysis is similar in essence to the methodology used in \cite{fan2020machine,wang2019transferable} connecting local structure and dynamics in several glass forming metals.
This provides a useful metric to measure structural (in)homogeneities within shear bands which also strongly couple to the glass ability to (de)localize strain.
The extent of bi-modality in the nonaffine displacement distributions is another signature of inherent heterogeneities in a glassy metal that may limit shear strain localization.
To our knowledge, this bi-modal behavior had not been previously reported in the literature.
Based on the radial distribution function data, sheared samples with uniform deformation patterns tend to have (relatively) abundant solid-like structures within flowing regions which may be also understood in terms of the heterogeneity notion.
The latter was further quantified topologically through the (effective) band width measurements which show meaningful variations with the rate of stress decay (as a bulk mechanical property) as well as intrinsic heterogeneities associated with the constituent elements' radii.   

\begin{figure}[b]
    \centering
    \begin{overpic}[width=0.5\textwidth]{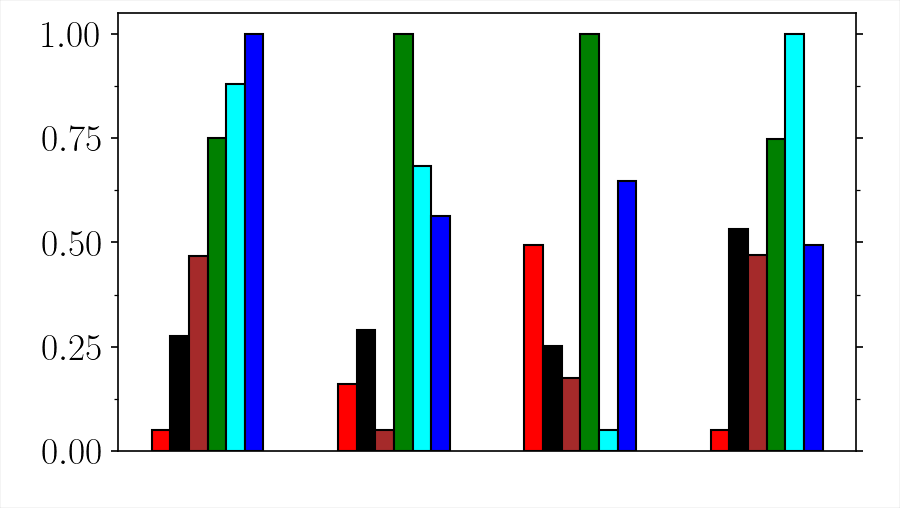}
        \Labelxy{22}{0}{0}{\large $h_\text{min}$}
        \Labelxy{42}{0}{0}{\large $\delta_a$}
        \Labelxy{62}{0}{0}{\large $w_\text{sb}$}
        \Labelxy{82}{0}{0}{\large $c^\infty_{XY}$}
        \Labelxy{0}{15}{90}{Scaled Quantities}
    \end{overpic}
    \caption{Softening modulus $h_\text{min}$, misfit parameter $\delta_a$, shear band width $w_\text{sb}$, and structure-dynamics correlations $c^\infty_{XY}$ corresponding to the FeNi(\protect\legRectTxt{0}{0}{black}), CoNiFe(\protect\legRectTxt{0}{0}{red}), CoNiCrFe(\protect\legRectTxt{0}{0}{darkspringgreen}), CoCrFeMn(\protect\legRectTxt{0}{0}{blue}), CoNiCrFeMn(\protect\legRectTxt{0}{0}{cyan}), Co$_5$Cr$_2$Fe$_{40}$Mn$_{27}$Ni$_{26}$ (\protect\legRectTxt{0}{0}{brown}) metallic glasses.}
    \label{fig:barchart}
\end{figure}

Figure~\ref{fig:barchart} provides a summary of key compositional/structural/mechanical metrics probed in this study.
The bar charts sort the element compositions for each index based on the ascending order in the softening modulus $h_\text{min}$ \add{(with the lowest $h_\text{min}$ marking the best glass)}.
Based on the observed trends in the data, $h_{\rm min}$ qualitatively correlates with $\delta_a$ but also, with the shear band width $w_\text{sb}$ and structure-dynamic correlations $c^{\infty}_{XY}$. 
\add{One may, however, infer stronger (positive) correlations between the studied order parameters within the scatter data as in Fig.~\ref{fig:cxyGamma}(b) and Fig.~\ref{fig:SoftModuAndWidth} owing to the fact that the bar charts illustrate \emph{mean} quantities associated with $h_\text{min}$, $w_\text{sb}$, and $c^{\infty}_{XY}$ over different realizations}.
\add{Our data indicate that good glasses ---with lowest $h_\text{min}$, $w_\text{sb}$, and $c^{\infty}_{XY}$--- statistically correlate with low misfits, opposite to the commonly-observed trend that the glass forming ability (in the thermodynamics sense) tends to improve with increasing $\delta_a$.
In other words, a glass that shears well is not a good glass former!
}

The observed correlations highlight the significance of chemical compositions and demand for a more systematic search strategy over a much broader composition space to design desired glasses. 
It is, therefore, essential to seek for a more descriptive and comprehensive set of relevant structural/physical/mechanical features and probe potential correlations with specific material properties, a training task amenable to materials informatics~\cite{frydrych2021materials}.   

The microstructural heterogeneity highlighted in our study must have a strong bearing on local elasticity and associated fluctuations in space \cite{wang2018spatial}.
The latter is a key element in STZ-based mesoscopic models which typically make no direct reference to atomic-scale disorder but instead take into consideration elastic heterogeneity as a main (often phenomenological) model ingredient.
These meso-scale treatments, however, largely ignore true microstructural origins associated with elastic inhomogeneity above some certain scales, which in turn impair predictive capacities of coarse-grained approaches. 
In this framework, a direct structural characterization of metallic glasses (via simulations and/or experiments) seems necessary in that local atomic configurations can provide insights on the nature of elastic heterogeneity and its multi scale notion in glassy materials.
These efforts should lead to significant improvements in existing coarse-grained models without compromising the true underlying physics.



\section{Conclusions}\label{sec:conclusions}
In summary, we have identified relevant structural features associated with differing degrees of strain localization observed in mechanically deforming model metallic glasses. 
We have based our approach on finding meaningful links between the micorstructure and atomic-scale dynamics in response to an external (shear) stress.
In particular, we established that \emph{\romannum{1}}) structural heterogeneity controls the glass quality described based on the degree of strain localization \emph{\romannum{2}}) the emerging short range order leads to a low susceptibility against localized rearrangements in good glasses enhancing their shear band instability \emph{\romannum{3}}) an abundance of short range ordering within shear zones in bad glasses increases structural inhomogeneity and local hardening and, therefore, their ability to delocalize strain.
Our results have important implications in terms of micro-structural tuning to optimize specific glass properties.

    




\bibliography{apssamp}

\end{document}